# High-Index Topological Insulator Resonant Nanostructures from Bismuth Selenide

*Sukanta Nandi[1,2], Shany Z. Cohen[1,2], Danveer Singh[1,2], Michal Poplinger[1,2], Pilkhaz Nanikashvili[1,2], Doron Naveh[1,2#]*

*and Tomer Lewi[1,2*]*

[1]Faculty of Engineering, Bar-Ilan University, Ramat Gan 5290002, Israel
[2]Institute of Nanotechnology and Advanced Materials, Bar-Ilan University Ramat Gan 52900, Israel
[*]Corresponding authors: [*]tomer.lewi@biu.ac.il, [#]doron.naveh@biu.ac.il

## Abstract

Topological insulators (TIs) are a class of materials characterized by an insulating bulk and high mobility topologically protected surface states, making them promising candidates for future optoelectronic and quantum devices. Although their electronic and transport properties have been extensively studied, their optical properties and prospective photonic capabilities have not been fully uncovered. Here, we use a combination of far-field and near-field nanoscale imaging and spectroscopy, to study CVD grown $Bi_2Se_3$ nanobeams (NBs). We first extract the mid-infrared (MIR) optical constants of $Bi_2Se_3$, revealing refractive index values as high as *n* ~6.4, and demonstrate that the NBs support Mie-resonances across the MIR. Local near-field reflection phase mapping reveals domains of various phase shifts, providing information on the local optical properties of the NBs. We experimentally measure up to $2\pi$ phase-shift across the resonance, in excellent agreement with FDTD simulations. This work highlights the potential of TI $Bi_2Se_3$ for quantum circuitry, non-linear generation, high-Q metaphotonics, and IR photodetection.

Topological insulators (TIs) constitute a distinct class of materials characterized by an energy gap in the bulk (insulator like), while also hosting time-reversal symmetry protected gapless surface states.[1–4] Fundamentally, TIs have strong spin-orbit coupling with the edge states allowing flow of unidirectional supercurrent. These novel quantum states have been theoretically predicted as well as experimentally demonstrated in a wide variety of materials and systems[5] such as CdTe/HgTe/CdTe quantum wells,[6] strained HgTe,[7] $Sb_2Te_3/Sb_2Te_{3-y}Se_y$,[8–11] Bismuth selenide ($Bi_2Se_3$),[9,12–15] $Bi_2Te_3$,[9,16] $Bi_{2-x}Sb_xTe_{3-y}Se_y$,[17] $MnBi_2Se_4$,[18] $ZrTe_5$[19] and more, with extensive research towards realizing novel electronic devices.[20,21] Recently, the unique properties of TIs have been explored in the context of electromagnetic and photonic systems.[22–30] Chalcogenide TIs are particularly interesting as they possess narrow bulk bandgap,[16,31] exhibit extremely high permittivity values,[32,33] and strong anisotropic behavior.[34] Recent demonstrations such as high harmonic generation,[29,35] plasmonic properties in the ultraviolet to visible range[36,37] and high mobility surface states in the mid-infrared (MIR) and THz spectral ranges,[13,38,39] exemplify the potential of TIs.[13,29,33,36,38–42]

Among these systems, $Bi_2Se_3$, a strong 3D TI with a large room temperature (RT) bulk energy gap of ~0.3-0.35 eV, and high permittivity values, is an interesting TI material for photonic applications.[9,15,43–48] Previous theoretical and experimental investigations have indicated that $Bi_2Se_3$ is a high-index material with the real part of the refractive index (RI) peaking around $n$~5.3-5.8 for the NIR regime, depending on the number of layers[46,49], while other studies have primarily focused on understanding the light-matter interaction in $Bi_2Se_3$ within the NIR to MIR spectral ranges.[50–52] However, experimentally measured fundamental properties such as permittivity, RI, and optical anisotropy in these TIs are still lacking in the infrared range, as well as the photonic capabilities of both bulk and nanostructured $Bi_2Se_3$. In this work we study the MIR optical properties of TI $Bi_2Se_3$ and utilize them to demonstrate high-index Mie-resonant nanostructures in chemical vapor deposition (CVD) grown $Bi_2Se_3$ nanobeams (NBs). Using infrared spectroscopy, along with multiple Lorentz-Drude oscillator modeling that accounts for both the bulk and the surface metallic states, we extract the optical constants of $Bi_2Se_3$ and demonstrate that this material system has RI values as high as $n$ ~6.4, with minimal losses across the MIR range. Polarized Fourier-transform infrared (FTIR) spectroscopy reveals that the NBs exhibit TE and TM Mie-resonant modes with the fundamental mode confined within nanobeams having cross-section sizes smaller than $\lambda/8$. We use scattering-type scanning near-field optical microscope (s-SNOM) to perform nanospectroscopy and extract the amplitude and phase of z-polarized resonant modes. Local near-field phase mapping reveals in-plane inhomogeneity allowing to estimate local variations in the complex optical constants and losses in the NBs. We experimentally demonstrate up to $2\pi$ phase shift across the resonance, for NBs placed on gold (Au) substrates. We highlight the role of loss in the phase pickup in the NB resonators and show that small changes in the imaginary part of the RI ($k$) lead to large variations in the phase. The results highlight the potential of $Bi_2Se_3$ as a versatile material for nanophotonic applications, by combining ultracompact components with very high-index values and low losses, that are coupled with topological quantum properties.

$Bi_2Se_3$ NBs were grown on sapphire substrate with typical substrate sizes of ~1 cm$^2$, using CVD. Details of the synthesis procedure can be found in the supporting information (SI). Structural characterization of the as-synthesized nanostructures shows that they were in the form of NBs of varying dimensions (SEM, Figure 1a), while elemental mapping using energy dispersive X-ray (EDAX) spectroscopy (Figure 1b), revealed that the NBs are indeed composed of pure Bismuth (Bi) and Selenium (Se) elements.

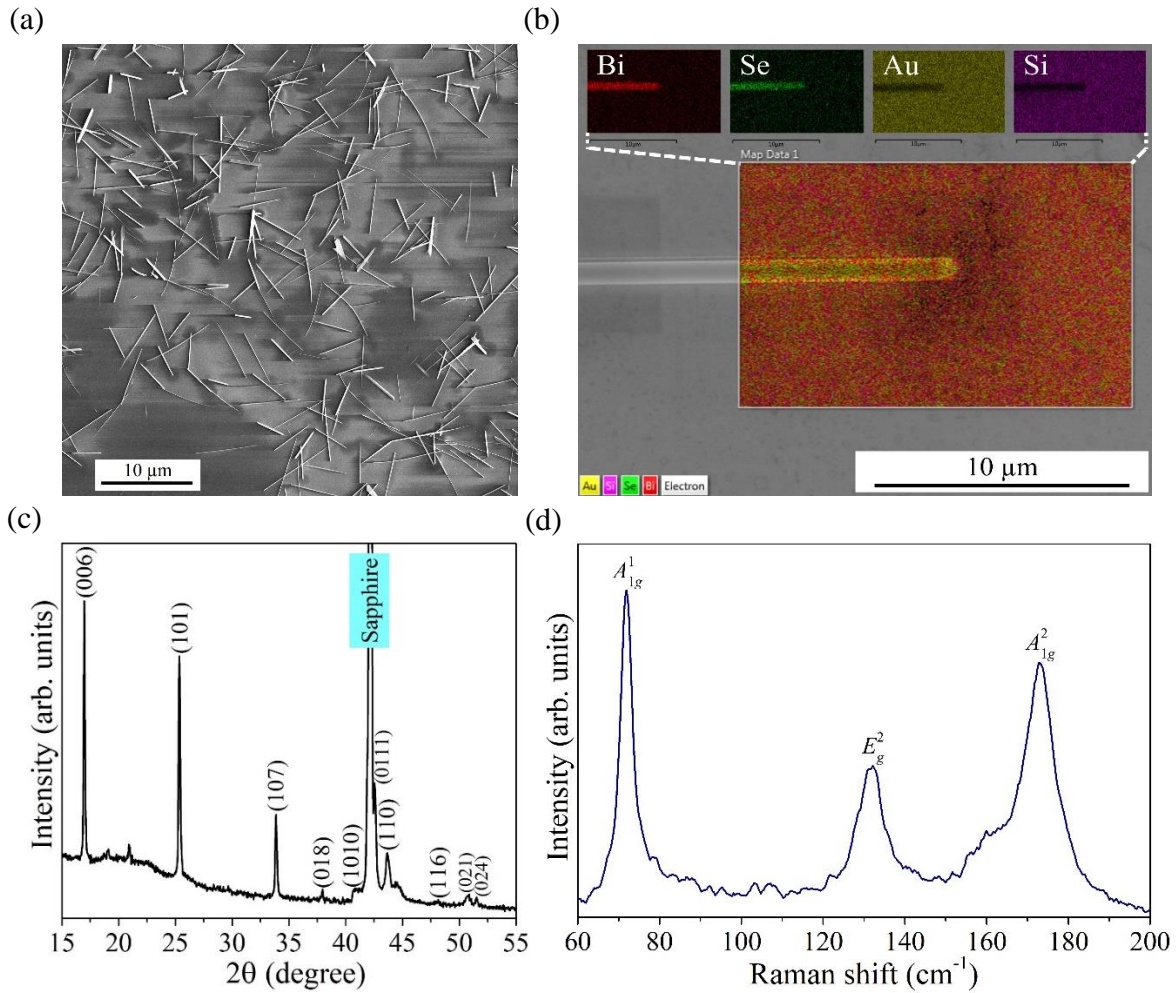

**Figure 1.** Material characterization of the CVD grown NBs. (a) SEM image of an ensemble of NBs and (b) that of a single NB together with its elemental mapping (placed on an Au substrate). (c) XRD and (d) Raman spectra of the NBs.

To understand the crystallographic properties and the chemical composition, we conducted X-ray diffraction (XRD) and single-particle Raman spectroscopy, as illustrated in Figure 1c and 1d, respectively. The indexed peaks in the XRD spectrum correspond to $Bi_2Se_3$ along with the background of sapphire (at ~42.16º, see SI for more details).[53–58] The Raman spectrum (Figure 1d) displays three characteristic peaks at ~72, ~132 and ~173 cm$^{-1}$, corresponding respectively to the $A_{1g}^1$, $E_g^2$ and $A_{1g}^2$ modes of Bi$_2$Se$_3$.[59–61]

SEM and atomic force microscopy (AFM) (see Figure S1 in SI) analysis indicated that the cross-section dimensions of the NBs have widths ($w$) of $w$~700 nm and heights ($h$) of approximately $h$~300 nm, with lengths varying between ~5 µm and ~30 µm. These observations suggest possible MIR Mie-scattering supported by the NBs.[62] However, a fundamental understanding of the scattering properties of these NBs requires precise knowledge of the Bi$_2$Se$_3$ optical constants. To obtain the optical constants, we performed spectroscopic reflection measurements (FTIR spectroscopy) on flakes of exfoliated Bi$_2$Se$_3$ of varying thicknesses, placed on Au substrate (see Figure S2 for a typical spectrum for a flake of thickness ~485 nm and corresponding AFM measurements in Figure S3). To analyze the experimental data and extract the dielectric function and complex RI of Bi$_2$Se$_3$, we used RefFIT program[63] in which a combination of Drude-Lorentz oscillators were used to model the material response. This fitted curve is based on modeling the material properties with two Lorentz oscillators for the bulk response and one Drude oscillator to account for the topological metallic surface, which was then used to extract the complex RI (Figure 2a). All model parameters are summarized in table 1 of the SI. Figure 2a presents the extracted real and imaginary parts of the complex RI ($\tilde{n} = n + ik$). The dispersion is characterized by very high index values peaking at $n$ ~6.42 for $\lambda$ ~2.56 µm, which can be attributed to the strong dielectric polarizability of the bulk. This is followed by a Drude-like descent for longer wavelengths, reflecting the combined contribution of both surface metallic states and bulk free carriers resulting from unintentional doping.[16,32,64–66] These latter effects are also responsible for the losses ($k$) in the system, which we further examine in the following sections.

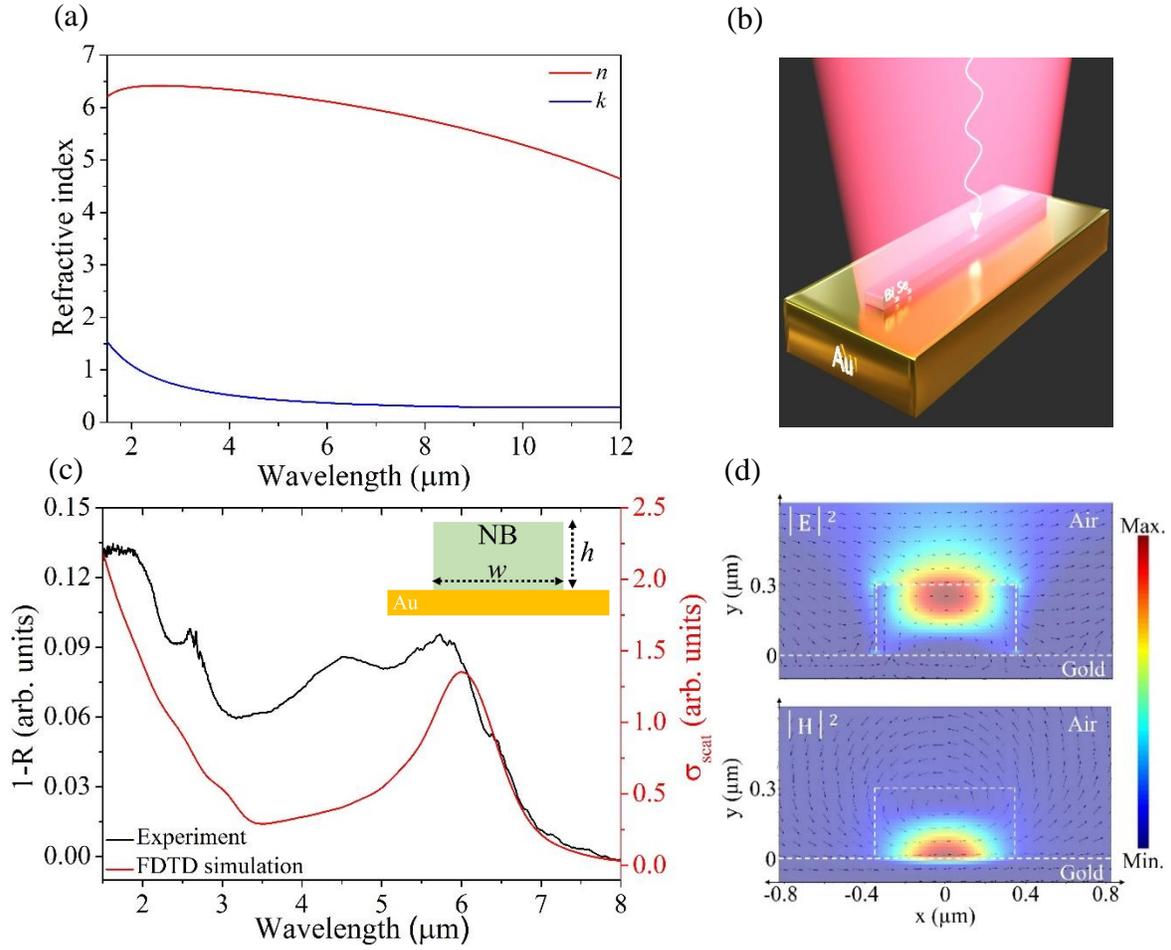

**Figure 2.** Optical properties of Bi$_2$Se$_3$ NBs. (a) Derived frequency-dependent complex optical constant of Bi$_2$Se$_3$. (b) Pictorial representation of Bi$_2$Se$_3$ NB on Au substrate illuminated by a linearly polarized light beam. (c) Comparison of experimentally measured extinction spectrum (black) and calculated FDTD scattering cross-section (red) from a single NB with *w = 0.7* µm & *h = 0.3* µm. Inset in (c) depicts the NB cross-section geometric parameters. (d) Cross-sectional view (XY) of the electric (*E*, top) and magnetic (*H*, bottom) field intensities together with their corresponding vector-field profiles at the resonance wavelength of λ=5.99 µm. White dashed outlines in the figure are guide to the eyes for locating the NB and its interfaces with air and Au substrate.

Nevertheless, the extremely high RI values reported here, originating from the ultra-high permittivity (presented in Figure S4 of SI), are the basis for the ultracompact nanophotonic components presented here.

To investigate the scattering characteristics of the CVD grown Bi$_2$Se$_3$ NBs we first used single particle spectroscopy using an FTIR (Nicolet, iS50R) spectrometer coupled to an infrared microscope (Nicolet, continuum infrared microscope).[67] Illustration of the illuminated NB on an Au substrate, is depicted in Figure 2b. The experimental spectra shown in Figure 2c, reveals the presence of two pronounced fundamental modes, at λ ~4.50 and ~5.73 µm, respectively. These results show that the NBs are ultracompact resonators with a dielectric like response within cross-section sizes as small as ~λ/8. Finite-Difference Time-Domain (FDTD) simulation (Details on the FDTD simulations can be found in the SI), incorporating the extracted optical constants (Figure 2a) and the NB dimensions, is in good agreement with the experimental results (Figure 2c). Figure 2d shows the electric (*E*, top) and magnetic (*H*, bottom) field profiles of the fundamental resonant mode.[68] The circulating *H*-field lines as seen in Figure 2d are a signature of a transverse magnetic (TM) resonance.

Figure 3a shows how the experimental unpolarized spectra of the NB can be decomposed into the two orthogonal resonant modes. This was done by inserting a polarizer into the incident beam path and selectively exciting either TE modes, where the *E*-field is perpendicular to the NB long-axis, or TM modes, where the *E*-field is parallel to the NB long-axis (see schematic at the top of Figure 3a). The unpolarized extinction spectrum is thus dominated by two modes, i.e., $TE_0$ (λ~ 4.50 µm) and $TM_1$ (λ ~5.73 µm), as seen in Figure 3a.[62] The geometric dispersion of the NBs is presented in Figures 3b and 3c, illustrating size dependencies of the resonant modes along a range of infrared wavelengths. In both cases the expected resonance red-shift with increasing size is observed (Figure S5, SI), allowing to engineer the scattering properties of the NBs. The very high index of $Bi_2Se_3$ enables the support of resonant modes in ultrathin NBs. Figure 3b shows that even a 100nm thick NB supports the $TM_1$ mode at λ~2.8µm, while the higher order mode has a modal cutoff for *h*~200nm at λ~1.5µm. When the height is fixed to *h*=300nm (Figure 3c), both the $TM_1$ and higher order modes are supported (within the width range, 0.3 µm < w <0.7 µm) throughout the studied spectral range 1.5-8.0 µm.

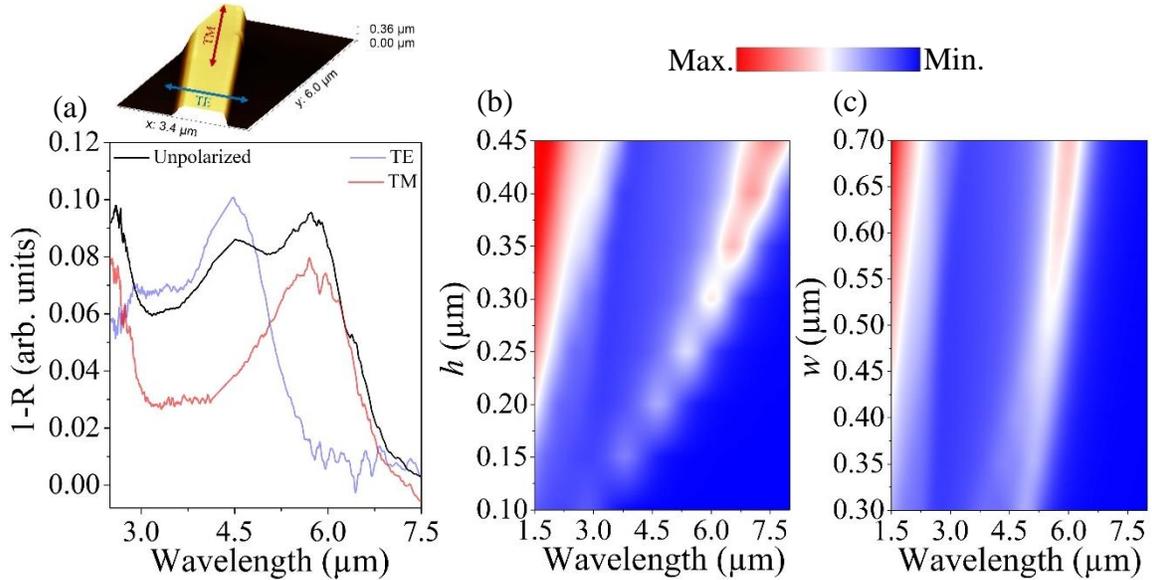

**Figure 3.** Polarized and size-dependent behavior of the NBs. (a) TE (blue) and TM (red) polarized scattering of the NB (*w = 0.7* µm & *h = 0.3* µm), along with the unpolarized (black) spectra. The top sketch illustrates the orientation of the TE and TM polarizations with respect to NB long-axis. (b, c) FDTD simulated modal evolution as a function of (b) height, *h* (for fixed *w = 0.7* µm) and (c) width, *w* (for fixed *h = 0.3* µm) variation. Color bar on the top indicates intensity of the resonant modes.

Altogether, this size-dependent modal behavior demonstrates how the very high index of $Bi_2Se_3$ facilitates resonant modes in subwavelength dimensions.

To gain more fundamental understating of the resonant nanostructures, we studied the near-field optical response and local properties of the $Bi_2Se_3$ NBs, using a dedicated setup consisting of s-SNOM, with an AFM tip as the near-field probe. The AFM probe and the optical apparatus are illustrated respectively in Figures 4a and 4b.

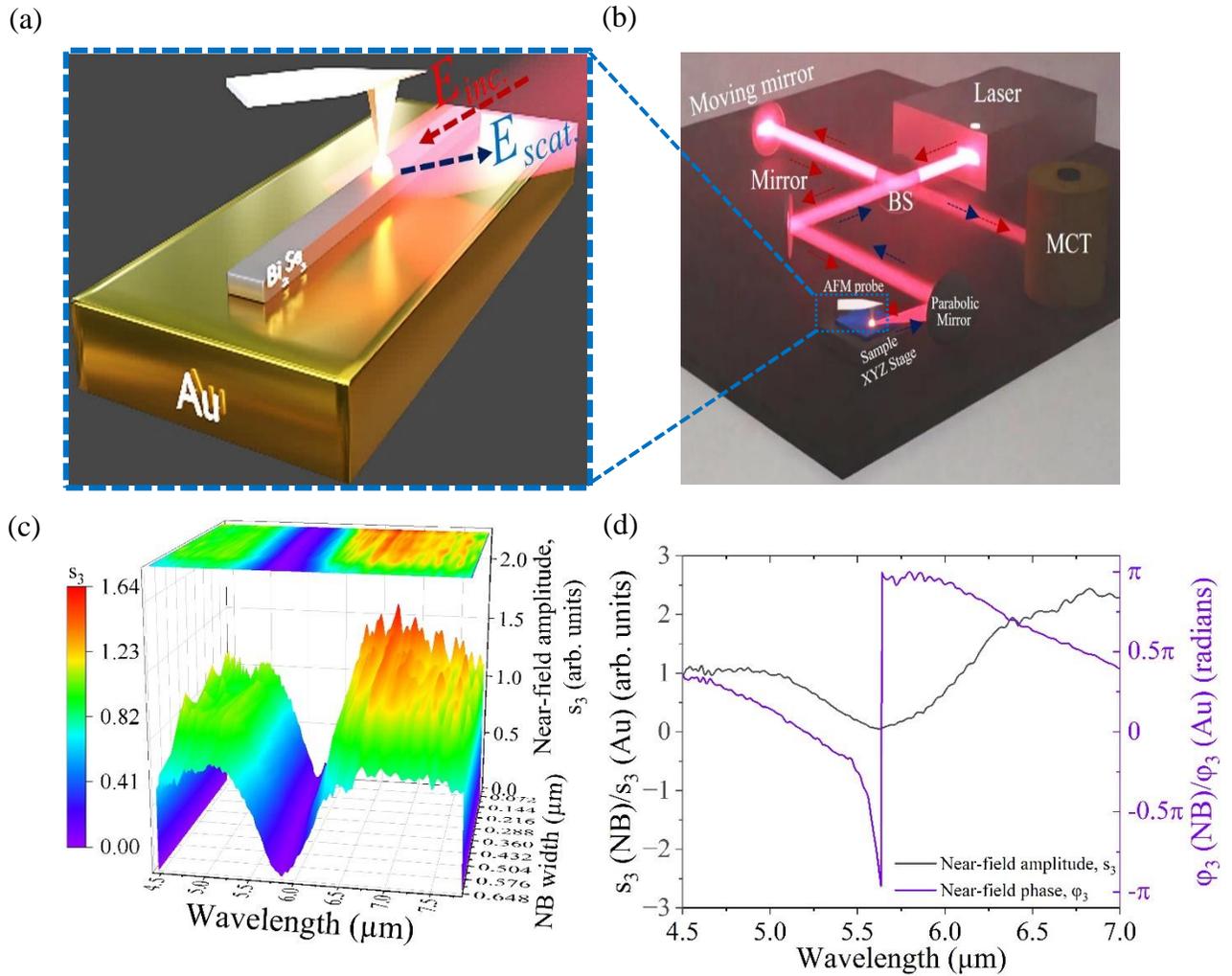

**Figure 4**. s-SNOM imaging and nanospectroscopy of $Bi_2Se_2$ NBs. (a) Illustration of near-field probing and consequent scattering from the $Bi_2Se_3$ NB placed on an Au substrate. (b) Schematic illustration of the s-SNOM setup with the heterodyne detection. BS refers to beam-splitter and MCT refers to Mercury-Cadmium-Telluride detector (c) Hyperspectral image, exhibiting mode evolution across the NB width. (d) Near-field scattered optical amplitude ($s_3$) and the corresponding phase ($\varphi_3$) from the NB.

s-SNOM based nanoscopy is a powerful technique for characterizing nanophotonic systems, material identification and probing exotic nanoscale and quantum phenomenon with 10-20 nm spatial resolution.[39,69–78] This technique allows to gather local information that is either lost or averaged out in the far-field response, as well to couple into excitations which are difficult to access from the far-field. In this study, we used a commercial s-SNOM (Neaspec, Attocube) configured with an asymmetric Michelson interferometer and a broadband MIR illuminating source (Figure 4b). More details of s-SNOM measurement and set-up can be found in the SI. Hyperspectral broadband nanospectrocopy measurements shown in Figure 4c, performed along the NB cross-section (*w*), exhibit a prominent resonant dip around λ ~5.65μm, which is consistently fixed along each scan position of the NB width. This measurement provided information on the nature of the spatial mode evolution, illustrating pinning of the resonant mode at the same spectral position (λ ~5.65μm) with a minuscule deviation of ~4% across a scan length of ~650 nm. These results suggest that the NB was grown with a high spatial uniformity. We also note that due to the nature of the z-oriented AFM probe, the near-field excitation (and collection) has a dominant *E*-field along the z-axis, which can explain the small discrepancy between the far-field $TM_1$ mode observed at λ ~5.73 μm and the mode excited by the near field probe (λ ~5.65μm). Local amplitude and phase scans, facilitated by the heterodyne detection configuration, reveal reflection phase shifts of up to 2π across the resonance wavelength, as seen in Figure 4d. Achieving 2π phase shifts (Figure 4d) is fundamental in nanophotonics and meta-optic design[68,79–83] and demonstrates the potential of $Bi_2Se_3$ in these fields. However, although the location of the resonance mode across the NB width and length, was consistently measured at λ ~5.65μm, mapping of the corresponding local reflection phase, revealed significant variations from the 2π phase shift presented in Figure 4d. In the following, we exemplify the capabilities of near-field microscopy to locally probe variations in the optical constants in different domains along the NB– variations which are otherwise averaged out in the far-field.

Figures 5a-c present experimental local phase measurements at three different domains along the NB, demonstrating the difference between three typically observed phases of φ~0.37π and φ ~0.72π radians (Figure 5a and 5b) and the maximum measured phase shift of φ

~2π radians (Figure 5c). In a typical scattering scenario, an optical phase shift of π radians is observed across a single scattering resonance.[85] However, proper engineering schemes can be adopted to obtain a full 2π phase shift.[86,87] One such approach, as implemented in this study (NB on an Au substate), is to place a resonator over a metal substrate, and if small or negligible ohmic losses exist in the system, this will result in a maximum of 2π phase shift,[87] due to the interference of the resonating mode with its mirror image in the substrate.[68,88] However, if ohmic losses exist in the system, then the maximum phase shift is not guaranteed. Physically, the imaginary part of the refractive index $k$, not only determines the losses in the material, but it also governs important resonator parameters such as light localization, quality-factor, and the optical phase shift. As we show here, even a small change in $k$, can lead to significant changes in the reflected phase from the NB resonator.

To understand the origin of these phase variations, we conducted a series of FDTD reflection phase simulations[84] (details on FDTD simulation can be found in the SI), for three different cases, as presented in Figures 5d-f. The reflected phase ($\varphi$) in Figure 5d was calculated using the experimentally extracted optical constants (Figure 2a), where $k$ ~0.39 (at wavelengths around the resonance wavelength) and is in excellent agreement with the experimentally observed reflected phase shift seen in Figure 5a. In Figures 5e and 5f we modified the optical constants to obtain slightly smaller $k$-values of $k=0.25$ (Figure 5e) and $k=0.20$ (Figure 5f), respectively, which provide excellent matches to the experimentally observed phase shifts (Figure 5b and 5c). However, the resulting reflection phase shifts are dramatically different in all the three cases. For the experimentally extracted index values ($k$ ~0.39), the corresponding phase shift is $\varphi$ ~0.37π, whereas it is doubled ($\varphi$ ~0.72π) for $k = 0.25$ and abruptly jumps to values near $\varphi$~2π for $k = 0.2$.

To make sure these local phase variations are not caused by random local imperfections or variations in NB dimensions, we measured the phase across the NB length. Figure 5g shows an experimentally obtained phase map for a continuous line scan of ~1 µm (uncertainty in height is ~1.7 nm) along the NB long axis. From this scan we observe that a relatively constant phase shift of ~$0.37\pi$ is obtained within this domain, in accordance with the simulated phase shift using the experimentally derived optical constants. This measurement further supports that the NB is homogeneous in terms of physical dimensions and geometry. Figure 5h on the other hand shows the calculated reflection phase for a continuous scan over the $k$-values (while the real part $n$ is kept fixed). The abrupt jump in the reflection phase shift around $k$ ~0.2 is evident. This point corresponds to the critical losses in the system in terms of phase pickup, and if these losses are kept below this value – a near $2\pi$ shift can be reached. SI Figure S6 shows that this specific critical $k_{cr}$ ~0.2 value is linked to the real part of the refractive index $n$, so that $(k/n)_{cr}$ is the normalized critical loss value, where in our case $n$~6 and $(k/n)_{cr}$~0.033.

Analyzing the experimental local phase observations in Figures 5a-c, in the context of these calculations (Figures 5d-g), we can infer that local variations in the optical constants of the NB exist, and more specifically in the imaginary RI, $k$-value. It should be noted that since we do not observe any notable shift in the resonance wavelength, variations in the real part of the index ($n$) cannot be the reason for this large change in the local phase, as we further demonstrate in the SI (Figures S7 and S8). Another factor that can critically cause these dramatic phase shift swings is the local variations in the NB geometric dimensions, i.e., $h$ and $w$. As we show in the SI (Figure S9), even large variations in the width or height of up to 33% in the NB, cannot account for the large observed phase shifts. After excluding the possibilities of spatial variations in the real part of the RI ($n$) and geometric heterogeneities as potential mechanisms for the experimentally observed large phase shifts, we can thus deduce that local variations in the imaginary part of the RI ($k$) are the most likely reason for these significant fluctuations. Possible physical mechanisms behind these local $k$ variations may include inhomogeneity in sample defects, bulk carrier concentration and the so-called "*aging-effect*" in TIs, an effect arising from changes in electronic structure of surface states with time.[64,89] In $Bi_2Se_3$, this effect is quite prominent and arises primarily from unintentional doping, native defects and vacancies.[64–66,90-92]

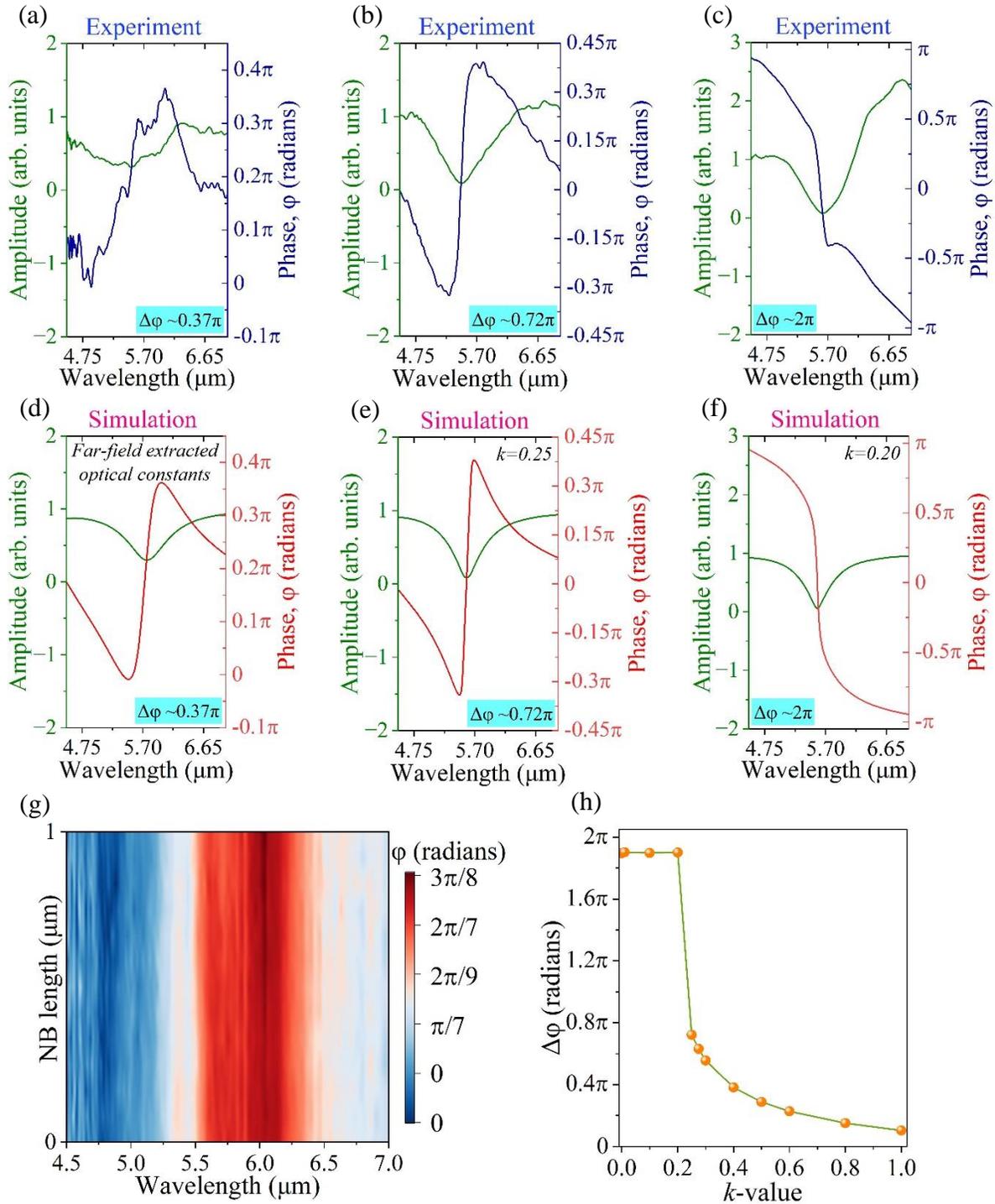

**Figure 5.** (a-c) Experimentally measured near-field amplitude and phase (φ) at different NB locations, resulting in a (a) φ~0.37π, (b) φ~0.72π and (c) φ~2π optical phase shifts. (d-f) Simulated near-field results with (d) extracted complex optical constants obtained from far-field measurements and varying *k*-values as, (e) *k*=0.25 and (f) *k*=0.20. (g) Optical phase map for a continuous line scan of ~1 µm along the NB long axis. (h) Detailed variation of near-field optical phase shift with varying *k*-values.

Finally, it is noteworthy to mention that NBs with larger dimensions, supporting resonances at longer wavelengths (λ ≥7.44 µm), would always exhibit a near-2π optical phase shift (see Figure S10 in SI). This response arises from the combined effect of larger NB dimensions (a higher Q-factor) and lower obtained *k*-values in the longwave IR regime (see Figure 2a).

Regardless, the experimentally measured 2π phase shifts observed in large domains of the samples suggest that $Bi_2Se_3$ within these areas exhibit low losses and demonstrate the power of TIs as a high index material for ultracompact nanophotonics. Furthermore, these local areas of minimal loss demonstrate the potential to obtain improved optical properties in TIs using advanced and optimized growth and fabrication techniques, potentially catalyzing further research into the underlying mechanisms. The near-field investigations conducted in this study illuminate a more intricate view of the spatially varying optical characteristics of the $Bi_2Se_3$ NB at the nanoscale, an information crucial for designing nanoscale photonic devices.

To summarize, in this work we demonstrate the synthesis of $Bi_2Se_3$ TI NBs of varying aspect ratios, using a CVD technique. Investigations on the optical properties revealed that the NBs are Mie-resonant nanostructures supporting both TE and TM polarized resonances, with very high RI values culminating in *n* ~6.42 and low extinction coefficient values, within a wide range of MIR wavelengths. Near field phase mapping probed by s-SNOM, revealed local reflection phase shifts of up to 2π, across the resonance. These local near field phase shift variations can be explained by small changes in the imaginary part of the index *k*, within different domains of the NB. The current study highlights the field of TI metaphotonics and demonstrates the potential of these high-index TI NBs for quantum circuitry, non-linear generation, high Q-factor metasurfaces, IR photodetection and many more.

## Acknowledgements

We would like to thank the Israel Science Foundation (ISF) for funding this work under grant no. 2110/19.

<u>**Supporting information**</u>

**High-Index Topological Insulator Resonant Nanostructures from Bismuth Selenide**

*Sukanta Nandi[1,2], Shany Z. Cohen[1,2], Danveer Singh[1,2], Michal Poplinger[1,2], Pilkhaz Nanikashvili[1,2], Doron Naveh[1,2#]*

*and Tomer Lewi[1,2*]*

[1]Faculty of Engineering, Bar-Ilan University, Ramat Gan 5290002, Israel
[2]Institute of Nanotechnology and Advanced Materials, Bar-Ilan University Ramat Gan 52900, Israel
[*]Corresponding authors: [*]tomer.lewi@biu.ac.il, [#]doron.naveh@biu.ac.il

**1. CVD growth of bismuth selenide nanobeams**

Nanobeams (NBs) of $Bi_2Se_3$ were synthesized by the technique of chemical vapor deposition (CVD) on a c-cut Sapphire [0001] substrate in a tube furnace inside a quartz tube. Before subjecting to CVD, the substrate surface was pre-treated in an ultrasonic bath with acetone and isopropyl alcohol for 5 mins to remove any contaminants. Organic leftovers from the surface were then cleaned using a solution of sulphuric acid and hydrogen peroxide ($H_2SO_4:H_2O_2$) in a ratio of 3:1 at 90 ºC for 15 mins, followed by washing in deionized water. For the CVD process, powder of $Bi_2Se_3$ (Alfa Aesar-99.999%) was used as the precursor. To begin the synthesis, an inert atmosphere inside the tube was maintained by flowing 25 sccm (standard cubic centimeter per minute) of nitrogen. The substrate was then maintained at a temperature of 400 ºC. Growth of the NBs was then carried out at 590 ºC (ramp rate of 20 ºC/min) for 3 hrs.

**2. XRD peaks of $Bi_2Se_3$ NB**

The peaks are observed at 2θ ~17°, 25.3°, 33.9°, 38°, 40.9°, 42.5°, 43.7°, 48.2°, 50.9° and 51.5° which correspond respectively to the (006), (101), (107), (018), (1010), (0111), (110), (116), (021) and (024) planes of $Bi_2Se_3$.

## 3. Atomic force microscopy of Bi$_2$Se$_3$ NB

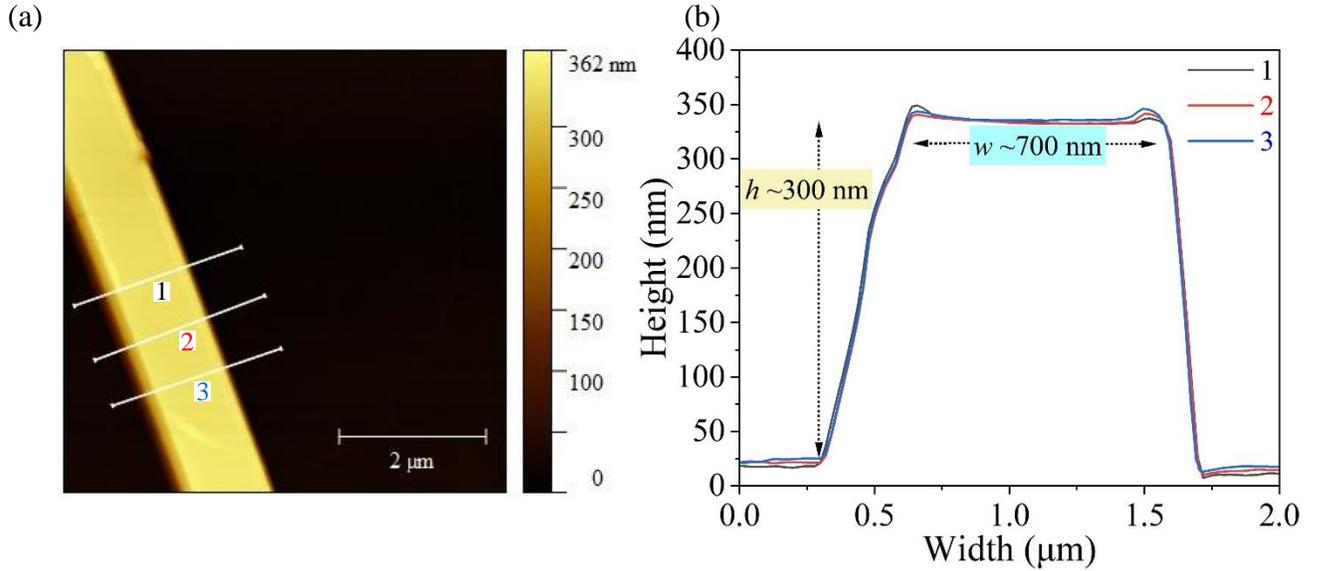

**Figure S1.** Atomic force microscopy (AFM) image of (a) Bi$_2$Se$_3$ NB and its (b) corresponding height ($h$) and width ($w$) extracted from three different locations, 1, 2, and 3.

## 4. Far-field spectroscopy

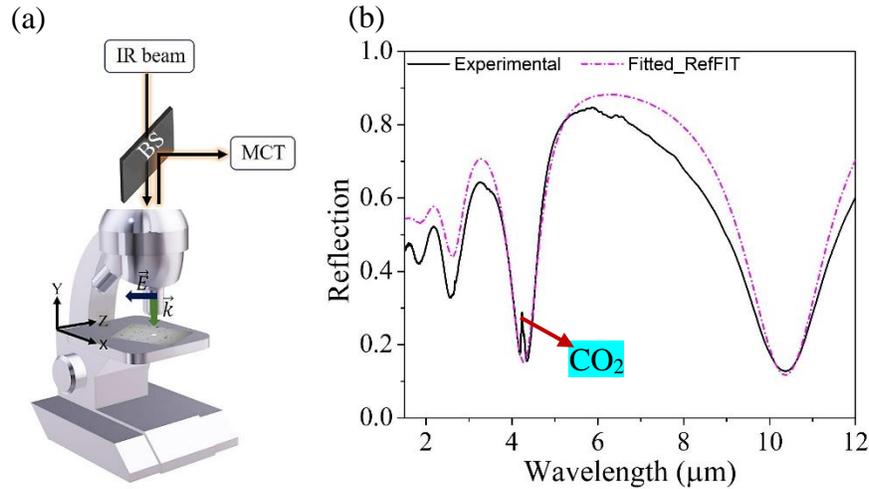

**Figure S2.** (a) Schematic illustration of far-field single particle infrared (IR) microspectroscopy (BS refers to beam-splitter & MCT to Mercury-Cadmium-Telluride detector). (b) Mid-IR reflectance spectra (experimental and fitted) of a Bi$_2$Se$_3$ flake of thickness ~485 nm. The sharp peak in the experimental spectra at ~4.2 μm is due to atmospheric $CO_2$ absorption.

## 5. AFM of Bi₂Se₃ flakes

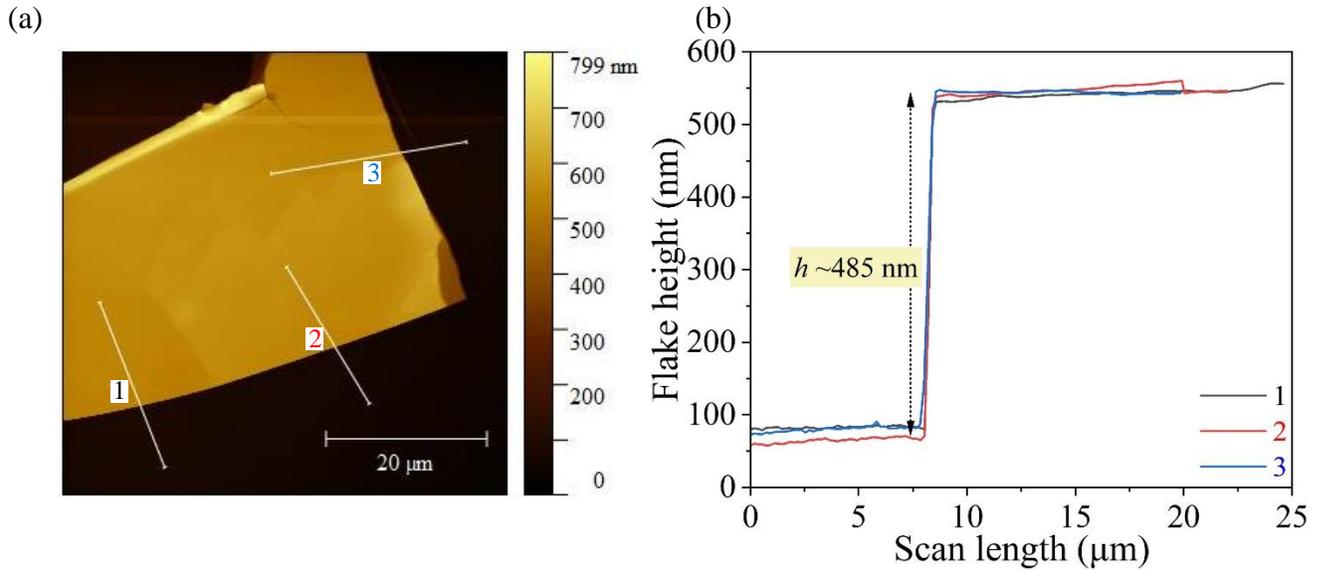

**Figure S3.** AFM image of (a) Bi$_2$Se$_3$ flake, and its (b) corresponding height extracted from three distinct locations, 1, 2, and 3.

## 6. Oscillator parameters

| Oscillator | $\omega_0$ (eV) | $\omega_p$ (eV) | $\gamma$ (eV) |
|---|---|---|---|
| **Drude** | - | 0.48 | 1.24E-6 |
| **Lorentz #1** | 0.5E-3 | 0.75 | 8.45 |
| **Lorentz #2** | 1.39 | 6.89 | 1.56 |

**Table 1.** The oscillators parameters $\omega_0$, $\omega_p$, and $\gamma$, are respectively the transverse frequency, plasma frequency and scattering rate. The high-frequency dielectric constant ($\varepsilon_\infty$) being 18.2.

## 7. Frequency-dependent permittivity of Bi₂Se₃

The spectral response for the variation of real ($\varepsilon_r$) and imaginary ($\varepsilon_i$) components of Bi$_2$Se$_3$ permittivity ($\varepsilon$) is shown in Figure S4.

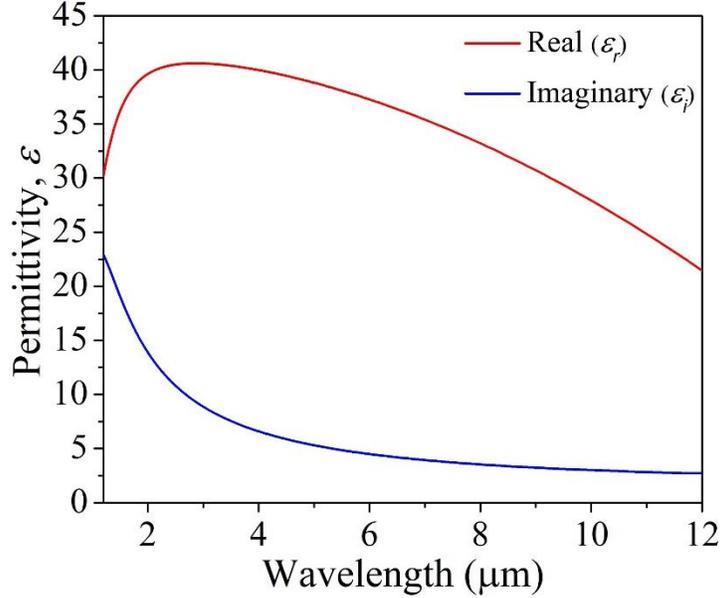

**Figure S4.** Spectral dependence of $Bi_2Se_3$ permittivity.

From Figure S4 we see that in the NIR spectral range (<1.5μm) the material is dominated by significant absorption due to the large value of the imaginary part of permittivity ($\varepsilon_i$) where its comparable with the real part ($\varepsilon_r$). For longer wavelengths ($\lambda$>2μm), the real part ($\varepsilon_r$) is extremely high with much smaller values of the imaginary part ($\varepsilon_i$), defining a dielectric like behavior for the $\lambda$ ~2 -12μm range.

## 8. Far-field simulations

The far-field scattering spectra of the single NB on Au substrate in the MIR optical regime was simulated using Ansys Lumerical's FDTD solver.[1] Considering the high aspect ratio of the NB (length/width ~40), we perform 2D simulations[2] to compute the simulated result, and then compare it with the experimental one. A simulation time of 1000 fs with auto non-uniform mesh under the mesh refinement of conformal variant 1 with a minimum mesh step of 0.00025 μm was used. Furthermore, for obtaining better resolution at sharp geometrical discontinuities, an additional uniform overriding mesh of 0.015 µm was applied. A perfectly matched layer (PML) boundary condition (BC) was then used along both the axes, being placed ~$2\lambda_{max}$ ($\lambda_{max}$ is the maximum wavelength of the sweep) away from the NB center. To illuminate the NB, a Total-Field scattered-field (TFSF) source with 45° polarization was used as the incident electromagnetic source. Scattering cross-section was then computed using the "cross_section" analysis object, being placed in between the TFSF and PML. Scattering results were then achieved using the as-obtained frequency-dependent complex optical constants of $Bi_2Se_3$ and Lumerical's in-built constant of Au (Gold)-Palik. Finally, the electric and magnetic field components of the scattered field were recorded using a Frequency-domain field and power 2D-Z normal monitor.

9. **Near-field simulations**

The near-field optical amplitude and phase was simulated using the "grating_s_parameters" analysis object of Ansys Lumerical's FDTD solver. This analysis object uses a plane-wave as the source of excitation. Further, to better mimic the experimental conditions, an incidence angle of 45° was set for the excitation using Broadband Fixed Angle Source Technique (BFAST) plane-wave type.[3] A 3D simulation was then set-up with a 2D monitor, placed 5 nm above the NB and parallel to the substrate plane (in this case a Perfect Electrical Conductor substrate, PEC) to record the frequency-dependent near-field amplitude and phase.[3] Two sets of BCs were used to evaluate the near-field characteristics. Along the injection direction (z-direction) PML BC was used, while along the other two axes, PML (overridden by BFAST plane wave), Bloch (overridden by BFAST plane wave) and Symmetric BCs were used. Similar computational results were obtained for both these cases. The simulation volume was ~$2 \times 30.63 \times 3.5$ µm³, and, to ensure sufficient decay of the resonance field, a simulation time of 5000 fs was set.

## 10. Static tunability via size engineering

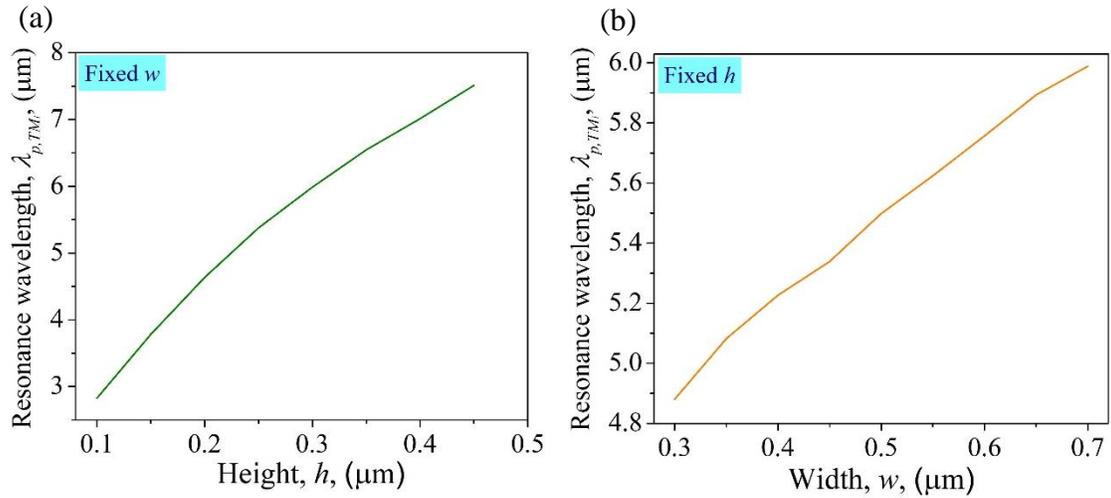

**Figure S5.** $TM_1$ mode evolution as a function of varying (a) height ($h$, fixed width) and (b) width ($w$, fixed height).

## 11. Near-field spectroscopy

Near-field measurements were performed by focusing a MIR beam both on the sample and the AFM tip using a parabolic mirror and then later collecting the backscattered light (Figure 4a and b, main text). The illuminated AFM probe creates a strong nano-focus on the sample and subsequently results in back-scattering ($E_{scat}$) (Figure 4a, main text). A pseudo-heterodyne detection mechanism is adopted in the system via the AFM cantilever being operated in the tapping mode. The backscattered light which then contains information of the local optical properties of the NB (amplitude and phase) is detected using a liquid nitrogen ($LN_2$) cooled MCT detector. The nanospectroscopy was performed with a broadband MIR laser (4.5−15 µm; FemtoFiber dichro midIR, Toptica) in an asymmetric interferometer configuration (Figure 4b, main text). Recorded interferogram is then Fourier transformed, and the scattered spectrum was then obtained by normalizing the NB spectrum with a reference spectrum, which in this case was Au spectrum (Figure 4d, main text).

## 12. Near-field amplitude and phase for varying *n* and fixed *k*

**Case I:** $k = k_{cr} = 0.2$

Figure S6 shows the near-field amplitude and phase as a function of varying *n* with fixed *k* as 0.2 (the critical value).

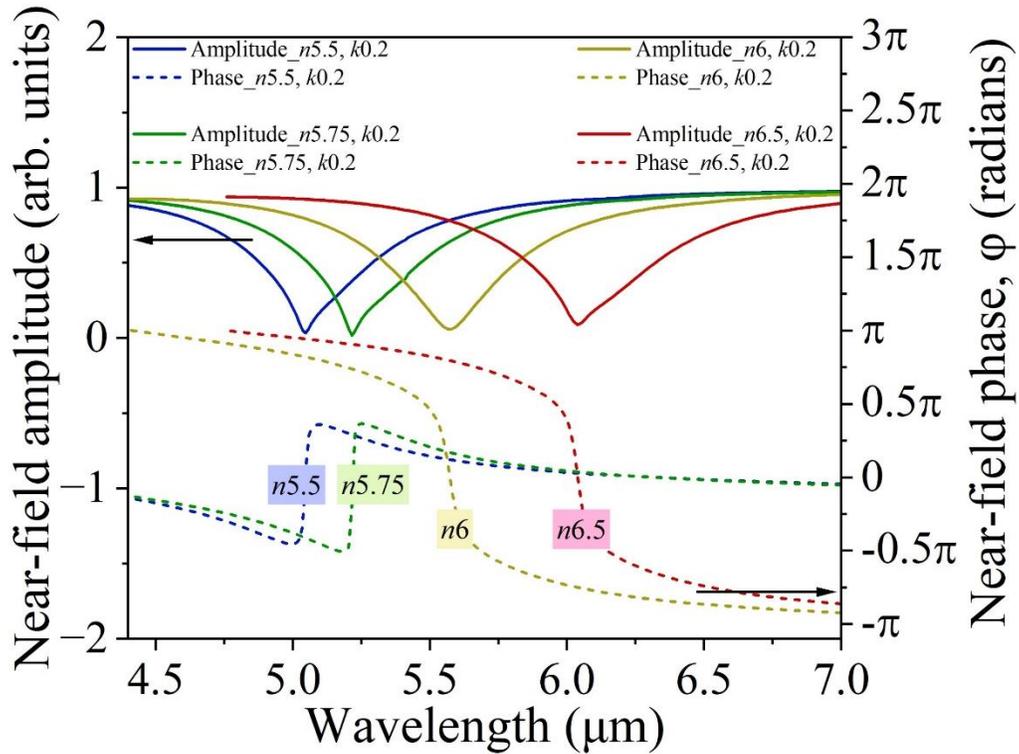

**Figure S6.** Near-field amplitude and phase for $k = 0.2$ and varying *n* as 5.5, 5.75, 6 and 6.5.

From the plot we see that increasing *n* from 5.5 to 6.5 red-shifts the resonance wavelength, as a typical Mie-resonance, however, the optical phase shift exhibits a large increase from $\Delta\varphi \sim \pi$ radians for $n<6$ to $\Delta\varphi \sim 2\pi$ for $n \geq 6$.

**Case II: $k = 0.39$**

Figure S7 presents the near-field amplitude and phase as a function of varying $n$ with fixed $k$ as 0.39 (obtained $k$-value around the resonance wavelength).

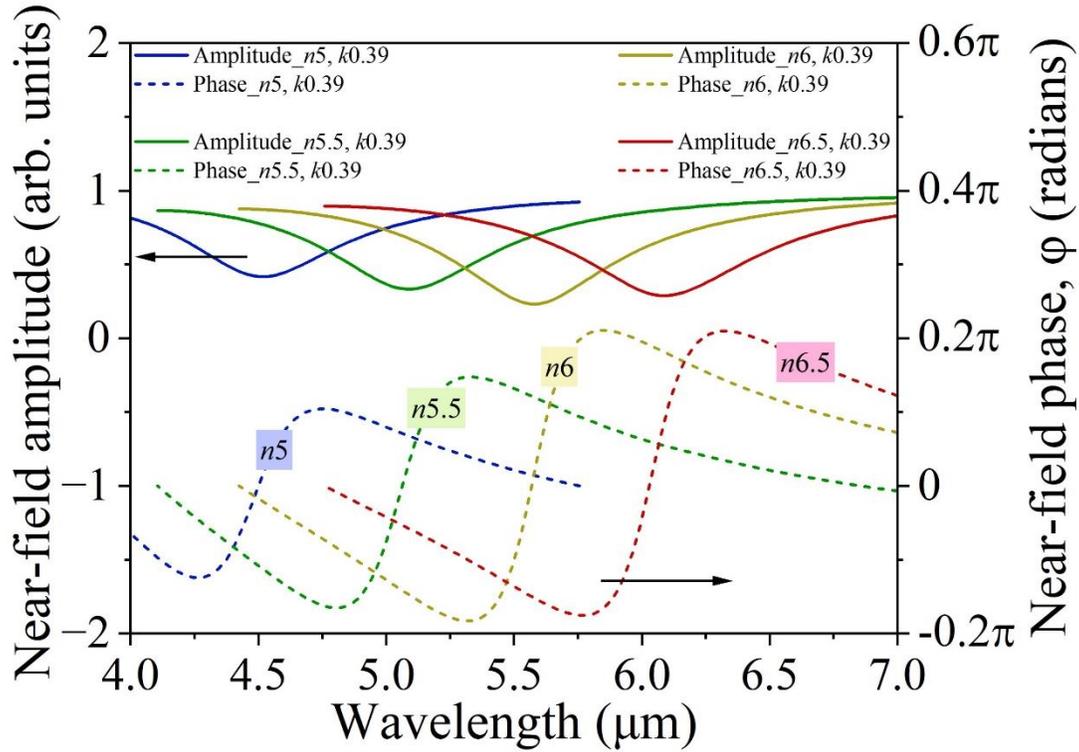

**Figure S7.** Near-field amplitude and phase for $k = 0.39$ and varying $n$ varying from 5.0-6.5.

From the plot we see that increasing $n$ from 5.0 to 6.5 red-shifts the resonance wavelength, as a typical Mie-resonance, however, the optical phase shift remains <0.5π, even for $n$ going as high as 6.5.

**Case III: $k = 0$**

Figure S8 shows the near-field amplitude and phase for the case of a purely dielectric NB ($k$ as 0). From the figure we see that, increasing $n$ from 5.5 to 6.5 red-shifts the resonance wavelength, as a typical Mie-resonance, however, the optical phase shift remains pinned at ~2π, thus suggesting null effect of *n* variation on the optical phase shift for the case of a purely dielectric system.

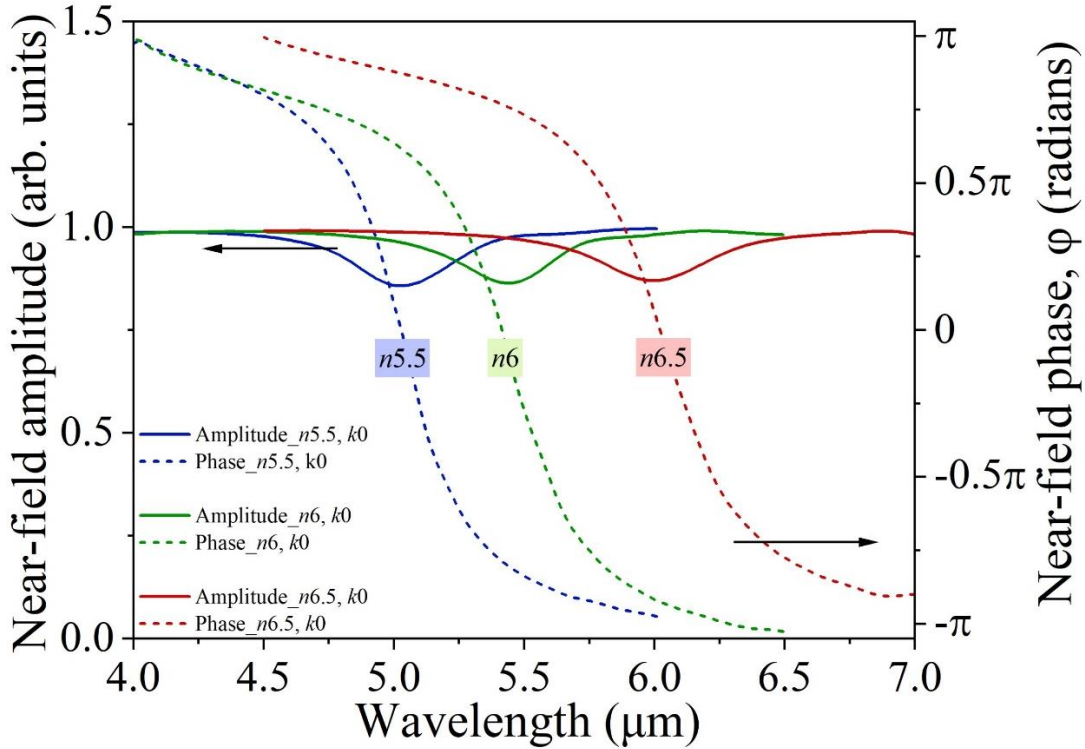

**Figure S8.** Near-field amplitude and phase for $k = 0$ and varying *n* as 5.5, 6 and 6.5.

### 13. Near-field amplitude and phase for varying NB geometric dimensions

Figure S9 presents the variation of near-field amplitude and phase as a function of increasing the NB dimensions along *h* and *w*. Due to z-polarized excitation, effect of *h* variation should be dominant for controlling the phase shift, thus, we calculated the reflection amplitude and phase for a NB with $h = 0.4$ µm and same *w* (0.7 µm). Increasing the NB *h* by 100 nm (~33%) resulted in red-shift of the resonance wavelength to ~6.3 µm and a phase shift of ~0.46π radians (Figure S9a). Alternatively, fixing the *h* at 0.4 µm and increasing the *w* by 300 nm (~43%) led to a resonance wavelength of ~7.13 µm and an optical phase shift of ~0.79π radians (Figure S9b).

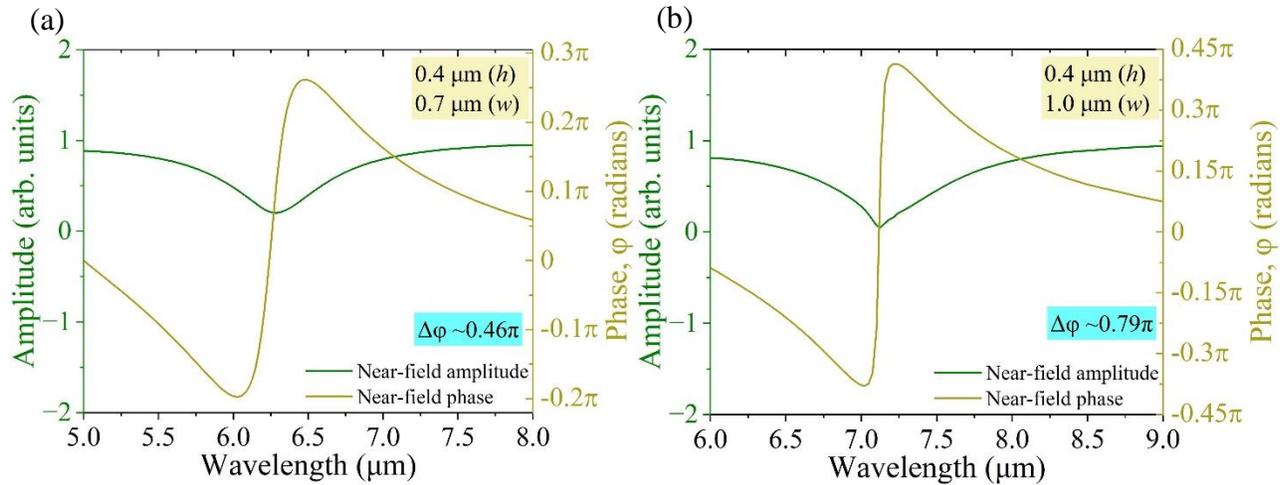

**Figure S9.** Near-field amplitude and phase for varying NB dimensions, (a) $h = 0.4$ µm, $w = 0.7$ µm, (b) $h = 0.4$ µm, $w = 1.0$ µm.

Increasing the NB dimensions red-shifts the resonance wavelength, as a typical Mie-resonance, with an achievable phase-shift of $\Delta\varphi < \pi$ for resonance wavelengths at ~7.13 µm and below.

## 14. Near-field amplitude and phase for NBs supporting resonances at longer IR wavelengths

Figure S10 presents the near-field amplitude and phase for NBs supporting resonances in the longer wavelength regime, always exhibiting a near ~$2\pi$ optical phase-shift.

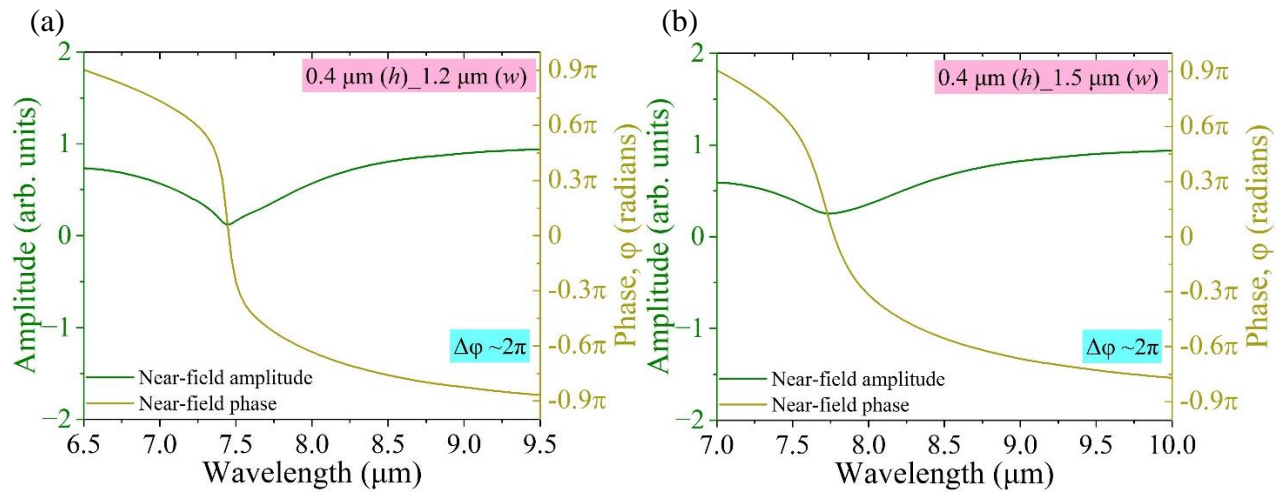

**Figure S10.** Near-field amplitude and phase for varying NB dimensions, (a) $h = 0.4$ µm, $w = 1.2$ µm, (b) $h = 0.4$ µm, $w = 1.5$ µm.